\documentclass[10pt]{article}
\usepackage{graphicx,floatflt,amssymb,epsf} 
\def\gtap{\raisebox{-.4ex}{\rlap{$\sim$}} \raisebox{.4ex}{$>$}} 
 
\def\e{{\cal{E}}}
\def\ehat{\hat{\cal{E}}}
\def\nn{{\cal{N}}_n} 
\def\n1{{\cal{N}}_{1i}} 

\begin{document} 
\begin{flushright} 
\texttt{hep-ph/0502031}\\ 
SINP/TNP/05-01\\ 
CU-PHYSICS/01-2005\\
\end{flushright} 
 
\vskip 30pt 
 
\begin{center} 
{\Large \bf Probing Universal Extra Dimension at the \\[1mm]
International Linear Collider} \\ 
\vspace*{1cm} 
\renewcommand{\thefootnote}{\fnsymbol{footnote}} 
{\large {\sf Gautam Bhattacharyya${}^1$}, {\sf Paramita Dey${}^1$}, 
{\sf Anirban Kundu${}^2$}, and {\sf Amitava Raychaudhuri${}^2$} 
} \\ 
\vspace{10pt} 
{\small ${}^{1)}$ {\em Saha Institute of Nuclear Physics, 
1/AF Bidhan Nagar, Kolkata 700064, India} \\  
   ${}^{2)}$ {\em Department of Physics, University of Calcutta, 
92 A.P.C. Road, Kolkata 700009, India}}  
 
\normalsize 
\end{center} 
 
\begin{abstract} 
In the context of an universal extra-dimensional scenario, we consider
production of the first Kaluza-Klein electron positron pair in an $e^+e^-$
collider as a case-study for the future International Linear Collider. The
Kaluza-Klein electron decays into a nearly degenerate Kaluza-Klein photon and
a standard electron, the former carrying away missing energy. The Kaluza-Klein
electron and photon states are heavy with their masses around the inverse
radius of compactification, and their splitting is controlled by radiative
corrections originating from bulk and brane-localised interactions. We look
for the signal event $e^+e^- +$ large missing energy for $\sqrt s = 1$ TeV and
observe that with a few hundred fb$^{-1}$ luminosity the signal will be
readily detectable over the standard model background. We comment on how this
signal may be distinguished from similar events from other new physics.
 
\vskip 5pt \noindent 
\texttt{PACS Nos:~ 12.60.-i, 14.60.Hi} \\ 
\texttt{Key Words:~~Universal Extra Dimension, 
International Linear Collider} 
\end{abstract}

\renewcommand{\thesection}{\Roman{section}} 
\setcounter{footnote}{0} 
\renewcommand{\thefootnote}{\arabic{footnote}}

{\bf Introduction}:~ If extra-dimensional models in a few hundred GeV
scale \cite{anto} are realised in Nature, one can not only undertake
their precision studies at the proposed International Linear Collider
(ILC) \cite{ilc} but also can distinguish them from other new
physics. In this paper, we consider such models with one extra
dimension having inverse radius of compactification in the range
$R^{-1} = 250-450$ GeV. We examine production of the first
Kaluza-Klein (KK) electron positron pair ($E_1^+ E_1^-$) in a linear
$e^+e^-$ collider operating at $\sqrt{s} = 1$ TeV.  The heavy modes
$E_1^\pm$ would decay into the standard (zero modes) $e^\pm$ and the
first KK photon ($\gamma_1$), the latter carrying away missing energy.
The splitting between $E_1^\pm$ and $\gamma_1$ comes from the bulk and
brane-localised radiative corrections. The cross section of the final
state $e^+e^-$ plus missing energy is quite large and the standard
model (SM) background is tractable, so that even with a one year run
of ILC at $\sqrt s$ = 1 TeV with approximately 300 fb$^{-1}$ enough
statistics would accumulate.  Forward-backward asymmetry of the final
state electron mildly depends on the initial polarisations. Even
though the mass spectrum of KK excitations of different SM particles
may resemble the supersymmetric pattern, angular distribution of the
final electrons can be used to discriminate the intermediate KK
electrons from selectrons or other new physics scalars.

{\bf Simplest universal extra dimension}:~ We consider the simplest
realisation of the universal extra dimension (UED) scenario in which
there is only one extra dimension which is accessed by all SM
particles \cite{acd}.  The extra dimension ($y$) is compactified on a
circle of radius $R$ along with a $Z_2$ orbifolding which renders all
matter and gauge fields, viewed from a 4 dimensional (4d) perspective,
depend on $y$ either as $\cos(ny/R)$ (even states) or $\sin(ny/R)$
(odd states), where $n$ is the KK index. The tree level mass of the
$n$th state of a particular field is given by $M_n^2 = M_0^2 +
n^2/R^2$, where $M_0$ is the zero mode mass of that field. Clearly,
excepting the top quark, Higgs, $W$, and $Z$, the KK states of all
other SM particles with the same $n$ are nearly mass degenerate at
$n/R$. Now, with all the fields propagating in the bulk, the momentum
along the fifth direction, quantised as $n/R$, remains a conserved
quantity. A closer scrutiny however reveals that a remnant $Z_2$
symmetry (different from the previous $Z_2$) in the effective 4d
Lagrangian dictates that what actually remains conserved is the KK
parity defined as $(-1)^n$. As a result, level mixings may occur which
admit even states mix only with even states, and odd with
odd. Therefore, (i) the lightest Kaluza-Klein particle (LKP) is
stable, and (ii) a single KK state (e.g. $n = 1$ state) cannot be
produced. These two criteria are reminiscent of supersymmetry with
conserved R-parity where the lightest supersymmetric particle (LSP) is
stable, and superparticles can only be pair produced. If produced, the
heavier KK modes can cascade decay to lighter ones, eventually to soft
SM particles plus LKP carrying away missing energy.  But low energy
constraints on the UED scenario from $g-2$ of the muon \cite{nath},
flavour changing neutral currents \cite{chk,buras,desh}, $Z \to
b\bar{b}$ decay \cite{santa}, the $\rho$ parameter \cite{acd}, other
electroweak precision tests \cite{ewued} and implications from hadron
collider studies \cite{collued}, all indicate that $R^{-1}~\gtap$ a
few hundred GeV.  As a result, even the second KK state having mass
$2/R$ will be beyond the pair-production reach of at least the first
phase of the planned linear collider. So, as mentioned in the
Introduction, we consider the production of first KK electron positron
pair and their subsequent decays into first KK photon plus the
standard leptons; the degeneracy between $E_1^\pm$ and $\gamma_1$
being lifted by radiative corrections which we shall briefly touch
upon below.

{\bf Radiative corrections and the spectrum:}~ 
Barring zero mode masses, the
degeneracy ($n/R$) at a given KK level is only a tree level result. Radiative
corrections lift this degeneracy \cite{cms1,pk,ggh,viq}. For intuitive
understanding, we consider the kinetic term of a scalar field as \cite{cms1}
$L_{\rm kin} = Z \partial_\mu \phi \partial^\mu \phi - Z_5 \partial_5 \phi
\partial^5 \phi ~~(\mu = 0,1,2,3)$, where $Z$ and $Z_5$ are renormalisation
constants. Recall, tree level KK masses ($M_n = n/R$) originate from the
kinetic term in the $y$-direction. If $Z = Z_5$, there is no correction to
those KK masses. But this equality is a consequence of Lorentz
invariance. When a direction is compactified, Lorentz invariance is lost, so
also is lost the equality between $Z$ and $Z_5$, leading to $\Delta M_n
\propto (Z-Z_5)$. One actually encounters two kinds of radiative
corrections. \\
(a) {\em Bulk corrections}: These corrections are finite. 
Moreover, they are
nonzero only for bosons. They arise when the internal loop lines wind around
the compactified direction, sensing that compactification has actually
occured, leading to the breaking of Lorentz invariance.  The correction to the
KK mass $M_n$ works out to be independent of $n$ and goes like $\Delta M_n^2
\propto \beta/16\pi^4 R^2$, where $\beta$ is a symbolic representation of the
collective beta function contributions of the gauge and matter KK fields
floating inside the loop.  Since the beta function contributions are different
for particles in different representation, the KK degeneracy is lifted.  One
can understand the decoupling of the correction as inverse power of $R$ by
noting that the $R \to \infty$ limit makes the fifth direction uncompactified
leading to exact Lorentz invariance. For the KK fermions this
correction is zero.  \\
(b) {\em Orbifold corrections}:~ Orbifolding additionally breaks 
translational
invariance in the fifth direction. The corrections to the KK masses arising
from interactions localized at the fixed points are not finite unlike the bulk
corrections. These are logarithmically divergent. These boundary terms can be
thought of as counterterms whose finite parts are completely undetermined. A
rather bold but predictive hypothesis is to assume that these corrections
vanish at the cutoff scale $\Lambda$.  Calculation shows that
the correction to $M_n$ does depend on $M_n$ in this case, and a generic
correction looks like $\Delta M_n \sim M_n (\beta/16\pi^2)
\ln(\Lambda^2/\mu^2)$, where $\mu$ is the low energy scale where we compute
these corrections.  The KK states are thus further split, this time with an
additional dependence on $\Lambda$.  

{\em Spectrum}:~ The mass spectra of the first excited electrons 
and the first excited $W^\pm, Z$ and photon for different choices of
$R$ and $\Lambda$ are displayed in Table 1. While the tree level KK mass is
given by $1/R$, the radiative corrections to them depend both on $R$ and
$\Lambda$ (for the exact expressions, see, e.g.,\cite{cms1}).
 
\begin{table} 
\begin{center} 
\begin{tabular}{|c|c|c|c|c|c|c|} 
\hline 
$R^{-1}$&$\Lambda R$&$M_{\ehat_1}$&$M_{\e_1}$&$M_{W_1}$& $M_{Z_1}$ & 
$M_{\gamma_1}$ \\ 
\hline 
250 &20 & 252.7 & 257.5 & 276.5 & 278.1 & 251.6 \\ 
\cline{2-7} 
      & 50 & 253.6 & 259.7 & 280.6 & 281.9 & 251.9 \\ \hline 
350 & 20 & 353.8 & 360.4 & 379.0 & 379.7 & 351.4 \\ \cline{2-7} 
      & 50 & 355.0 & 363.6 & 384.9 & 385.4 & 351.5 \\ \hline 
450 & 20 & 454.9 & 463.4 & 482.9 & 483.3 & 451.1 \\ \cline{2-7} 
      & 50 & 456.4 & 467.5 & 490.6 & 490.8 & 451.1 \\ \hline 
\end{tabular} 
\caption[]{KK masses ($n=1$) for different cases: excited electrons in
SU(2) singlet and doublet representations, excited charged and neutral
gauge bosons, respectively. All mass scales are in GeV.} 
\end{center} 
\end{table}

{\bf Production and decay modes of KK leptons}:~ The SU(2) doublet KK states
appear with both left and right chiralities as ${\mathcal{L}}_{L,R}$, where
${\mathcal{L}} = (\nn, \e_n)^T$, so do the SU(2) singlets ${\ehat}_{L,R}$. All
these states for $n=1$ will be pair produced at the foreseeable collider
energy. As noted in Table 1, the orbifold corrections create enough mass
splitting between these states and $\gamma_1$ (dominantly $B_1$) allowing the
former to decay within the detector to $e^\pm~ +$ missing energy which
constitute our signal. Below we denote ${\e}^\pm_1$ and ${\ehat}^\pm_1$
collectively by $E_1^\pm$.

Now we consider the pair production $e^+e^- \to E_1^+ E_1^-$ for
different polarisations of the incident beams.  The interaction
proceeds through $s$- and $t$-channel graphs.  The $s$-channel
processes are mediated by $\gamma$ and $Z$. The $t$-channel processes
proceed through $\gamma_1/Z_1$ gauge bosons and $\gamma^5_1/Z^5_1$
scalars (fifth components of 5d neutral gauge bosons).  $E_1$ decays
into $e$ and $\gamma_1$.  The splitting between $E_1$ and $\gamma_1$
masses is sufficient for the decay to occur well within the detector
with a 100\% branching ratio (BR). It may be possible to observe even
a displaced vertex (e.g., $\ehat_1$ decays, for $R^{-1} =$ 250
GeV). So in the final state we have $e^+e^-~+~ 2\gamma_1$ ($\equiv$
missing energy).
 
The same final state can be obtained from $e^+e^- \to W_1^+W_1^-$ as
well.  Again, the interaction proceeds through $\gamma$ and $Z$
mediated $s$-channel graphs, and ${\cal{N}}_{1L}$ mediated $t$-channel
graphs. Given the splittings (Table 1), $W_1^\pm$ can decay into
$e_i^\pm$ and $\n1$, as well as into $E^\pm_{1i}$ and $\nu_i$, where
$i=$ 1 to 3 is the flavour index. While $\n1$ escapes undetected,
$E^\pm_{1i}$ decays into $e_i^\pm$ and $\gamma_1$. So, if we tag only
electron flavours (plus missing energy) in the final state, the
$e^+e^- \to W_1^+W_1^-$ cross section, which is in the same ball-park
as the $e^+e^- \to E_1^+ E_1^-$ cross section, should be multiplied by
a BR of $\sim$ 1/9. Numerically, therefore, this channel is not
significant. Even more insignificant contribution would come from
$(W^5_1)^\pm$ scalar (fifth component of 5d charged gauge bosons) pair
production.
 
{\bf SM background}:~ The main background comes from $\gamma^*
\gamma^* \to e^+e^-$ events, where $\gamma^*$s originate from the
initial electron-positron pair while the latter go undetected down the
beam pipe \cite{colorado}. The $\gamma^* \gamma^*$ production cross
section is $\sim 10^4$ pb. About half of these events results in final
state $e^+e^-$ pair as visible particles.

The background $e^+e^-$ pairs are usually quite soft and coplanar with
the beam axis \cite{peskin}. An acoplanarity cut significantly removes
this background. Such a cut, we have checked, does not appreciably
reduce our signal. For example, excluding events which deviate from
coplanarity within 40 mrad reduces only 7\% of the signal cross
section. In fact, current designs of LC envisage very forward
detectors to specifically capture the `would-be-lost' $e^+e^-$ pairs
down the beam pipe\footnote{ To counter the two-photon background we
may also advocate the following strategy.  Instead of eliminating the
background, we calculate the number of $e^+e^-$ events originating
from two-photon production. For this we first count the number of
$\mu^+\mu^-$ plus missing energy events. The number of such events
coming from the decay of KK muons, we have checked for $1/R \sim 250 -
300$ GeV, would be rather small, about a factor of 1/20 compared to
the number of $e^+e^-$ plus missing energy events, due to strong
$s$-channel suppression. So most of the observed $\mu^+\mu^-$ events
would have sprung from $\gamma^*\gamma^*$. Thus the muon events serve
as a normalisation to count the $e^+e^-$ plus missing energy events
originating from the two-photon background. Our signal events should
be recognized as those which are in excess of that. Based on the
estimates of two-photon events given by the Colorado group
\cite{colorado}, we have checked that for an integrated luminosity of
300 ${\rm fb}^{-1}$ the signal events would be about ten times larger
than the square-root of background.}.

Numerically less significant backgrounds would come from $e^+e^- \to
W^+W^-$, $e\nu W$, $e^+e^-Z$, followed by the appropriate leptonic
decays of the $W$ and $Z$.

{\bf Collider parameters}:~ The study is performed in the context of
the ILC \cite{ilc}, running at $\sqrt{s}=1$ TeV (upgraded option), and
with a polarisation efficiency of 80\% for $e^-$ and 50\% for $e^+$
beams.  We impose kinematic cuts on the lower and upper energies of
the final state charged leptons as 0.5 and 20 GeV respectively.  While
the lower cut is a requirement for minimum energy resolution for
identification, the (upper) hardness cut eliminates most of the SM
background.  We also employ a rapidity cut admitting only those final
state electrons which are away from beam pipe by more than $15^\circ$.
 
{\bf Cross sections}:~ The cross section for $e^+e^-$ plus missing
energy final state has been plotted in Fig.~1. We have neglected the
events coming from excited $W$ decay. Notice that varying the beam
polarisations does create a detectable difference in the cross
section, nevertheless, there is no special gain for any particular
choice: for left-polarised $e^-$ beam, both $B_1$ and $W^3_1$
contribute, whereas for the right-polarised $e^-$ beam, only $B_1$
contributes but with an enhanced coupling.  The cross section enhances
as we increase $\Lambda R$ from 2 to 20; this is due to the change in
$\theta_{W1}$ (the weak angle for $n=1$ KK gauge bosons).  Further
increase of $\Lambda R$ does not change the cross section; a
saturation point is reached. Additionally, the kinematic cuts tend to
reduce the cross section which is why the curve for $\Lambda R = 50$
lies between the ones for $\Lambda R = 2$ and 20.
 
{\bf Forward-backward (FB) asymmetries}:~ The FB asymmetries of the
final state electrons, defined as $A_{\rm FB} = (\sigma_{\rm F} -
\sigma_{\rm B})/ (\sigma_{\rm F} + \sigma_{\rm B})$, are plotted in
Fig.~2 for different values of $\Lambda R$. The reason as to why it
falls with increasing $1/R$ is as follows. The first-stage process
$e^+e^-\to E_1^+E_1^-$ is forward-peaked, and for smaller $1/R$,
i.e. lighter KK electrons, the final state $e^\pm$ are boosted more
along the direction of the parent $E^\pm$. As $1/R$, or equivalently
the KK mass, increases the boost drops and the distribution tends to
lose its original forward-peaked nature.  Polarisation of the beams
does not appear to have a marked advantage. A point to note is that
the electrons coming from two-photon background will be FB symmetric.

{\bf Discriminating UED from other new physics}:~ 
It is not our purpose in this brief note to discuss at depth any
specific version of new physics model and its possible discrimination
from UED. Still, for illustrative purposes, we recall that the
spectrum of KK excitations for a given level (here $n = 1$) may be
reminiscent of a possible supersymmetry spectrum \cite{cms2}, where
the KK parity is `like' the R-parity.  Even in a situation when the
LSP weighs above 250 GeV and conspires to be almost degenerate with
the selectron, it is possible to discriminate a KK electron decaying
into the KK photon (LKP) from a selectron decaying into a neutralino
LSP by studying the angular distribution pattern of the final state
electron. We demonstrate this with a simple toy example. Compare the
pair production of (a) generic heavy fermions and (b) generic heavy
scalars in an $e^+e^-$ collider in a toy scenario. Assume $\sqrt{s}
\gg m$, where $m$ is the mass of the heavy lepton/scalar, so that only
the $t$-channel diagrams, with just a heavy gauge boson in case (a)
and a heavy fermion in case (b) as propagators, are numerically
dominant (we assume this only for the ease of analytic
comparison). The heavy states are produced with sufficient boost,
therefore the tagged leptons they decay into have roughly the same
angular distributions as them. Take the mass of the $t$-channel
propagator in either case to be about the same as the mass of the
heavy lepton/scalar as $m = 250$ GeV. For these choices, the ratio of
$d\sigma/d\cos\theta$ (case (a)/ case (b)) is observed to be $(3.8 +
1.3 \cos\theta + 0.6 \cos^2\theta)/\sin^2\theta$, clearly indicating
that the two cases can be easily distinguished from their angular
distributions. Moreover, the UED cross section is found to be a factor
of 4 to 5 larger than the scalar production cross section for similar
couplings and other parameters. For selectron production, indeed one
must take the detailed neutralino structure and the exact couplings,
but the basic arguments that we advanced for distinguishing scalar-
from the fermion-productions at the primary vertex using the toy model
would still hold.

{\bf Comparison with the CLIC Working Group study}:~ 
Our analysis is complementary to that in the CLIC multi-TeV linear
collider study report \cite{clic}. While we have electrons in the
final state, the study in \cite{clic} involves muons.  Clearly, the
angular distribution in our case is dominated by $t$-channel diagrams,
while the process studied in \cite{clic} proceeds only through
$s$-channel graphs.  Due to the inherently forward-peaked nature of
the $t$-channel diagrams, we obtain a significantly larger FB
asymmetry. Unlike in \cite{clic}, we have neither included the initial
state radiation effect nor incorporated detector simulation.

{\bf LHC/ILC synergy}:~ Extensive studies have been carried out
\cite{lhc-lc} addressing the physics interplay between the LHC and the
ILC, in particular, how the results obtained at one machine would
influence the way analyses would be carried out at the other. While
LHC may serve as a discovery machine, precision measurements of the
masses, decay widths, mixing angles, etc., of the discovered particles
can be carried out at the ILC. To illustrate this with an example, let
us consider a selectron weighing around 200 GeV. The analysis
\cite{m-selectron} shows that while the uncertainty in its mass
determination is around 5 GeV at the LHC, with inputs from the ILC the
uncertainty can be brought down to about 0.2 GeV due to a
significantly better edge analysis in the clean ILC environment.
Similar precisions may be expected for the masses of the KK electrons
as well for a comparable cross-section.  However, if $R^{-1}$ is large
and the cross section goes down by about a factor of 50-100, the
sensitivity will also go down scaling inversely as the square root of
the number of events. (Beam polarisation will not be of much help, as
can be seen from Fig.\ 1). Another point that may play a significant
role in determining the masses is the softness of final-state
electrons. We have applied adequate softness cuts to remove very soft
electrons.  One needs a detailed analysis to determine the exact
accuracies at different benchmark points, but roughly the accuracy of
determining the KK masses at the ILC should be of the order of 1 GeV
or even better.  Even if the KK electrons are first observed at the
LHC, their spin assignments might not be possible. This particular
issue, i.e. whether UED states can be distinguished from the
supersymmetric states at the LHC, has been studied in
\cite{webber}. Considering the decay chain in which a KK quark (or, a
squark) disintegrates into a quark, a lepton-antilepton pair and
missing energy, and looking at the spin correlations of the emitted
quark with one of the leptons, the authors of \cite{webber} conclude
that for a quasi-degenerate (UED like) spectrum, spin assignments of
the discovered particles could hardly be efficiently done: the best
discriminator of UED from supersymmetry in that case would be a
significantly larger production cross sections for the UED particles
than those of the supersymmetric ones. On the contrary, if the
observed spectrum is hierarchical (e.g. a supersymmetric type), the
prospects of observing spin correlations would be better.  However,
the ILC would provide a better environment for doing spin studies.
The clinching evidence of UED would of course be the discovery of the
$n = 2$ KK modes. While the $Z_2$ peak can be discovered at the LHC
through some hadronically quiet channels, $\gamma_2$ will be hard to
detect at the LHC because it immediately decays into two jets which
will be swamped by the QCD background \cite{biplob}. Turning our
attention now to the ILC, given its proposed energy reach, these
states will too heavy to be pair produced, but as shown in
\cite{biplob}, single resonant productions of $Z_2$ and $\gamma_2$,
despite suppression from KK number violating couplings, will have
sizable cross sections. Precision measurements of their peak positions
and widths at the ILC will enable one to extract $R$ and $\Lambda$.

{\bf Conclusion}:~ We have shown that the ILC may have a significant
role in not only detecting the presence of few-hundred-GeV-size extra
dimensions but also discriminating it from other new physics options,
like supersymmetry. Even if the KK modes are first observed at the
LHC, one needs the ILC for their proper identification through
precision measurements of the masses, couplings and spin correllations.
The physics interplay of the LHC and the ILC will be quite important
in this context.

\vskip 10pt

\centerline{\bf{Acknowledgements}} We thank H.C. Cheng for clarifying
to us some aspects of orbifold radiative corrections. We acknowledge
very fruitful correspondences with M.E. Peskin on the two-photon
background. We also thank J. Kalinowski for presenting a preliminary
version of this work in ICHEP 2004, Beijing \cite{kali}.  Thanks are
also due to S. Dutta and J.P. Saha, who were involved in the earlier
stages of this work. Stimulating discussions with the participants
of the Study Group on Extra Dimensions at LHC, held at HRI, Allahabad,
are also acknowledged. G.B., P.D., and A.R. acknowledge hospitality at
Abdus Salam ICTP, Trieste, while G.B. also acknowledges hospitality at
LPT, Orsay, and Theory Division, CERN, at different stages of the
work.  G.B. and A.R. were supported, in part, by the DST, India,
project number SP/S2/K-10/2001. AK was supported by DST, Govt.\ of
India, through the project SR/S2/HEP-15/2003.

\newpage 

\begin{figure} 
\vspace{-10pt}
\centerline{\hspace{-3.3mm} 
\rotatebox{-90}{\epsfxsize=8cm\epsfbox{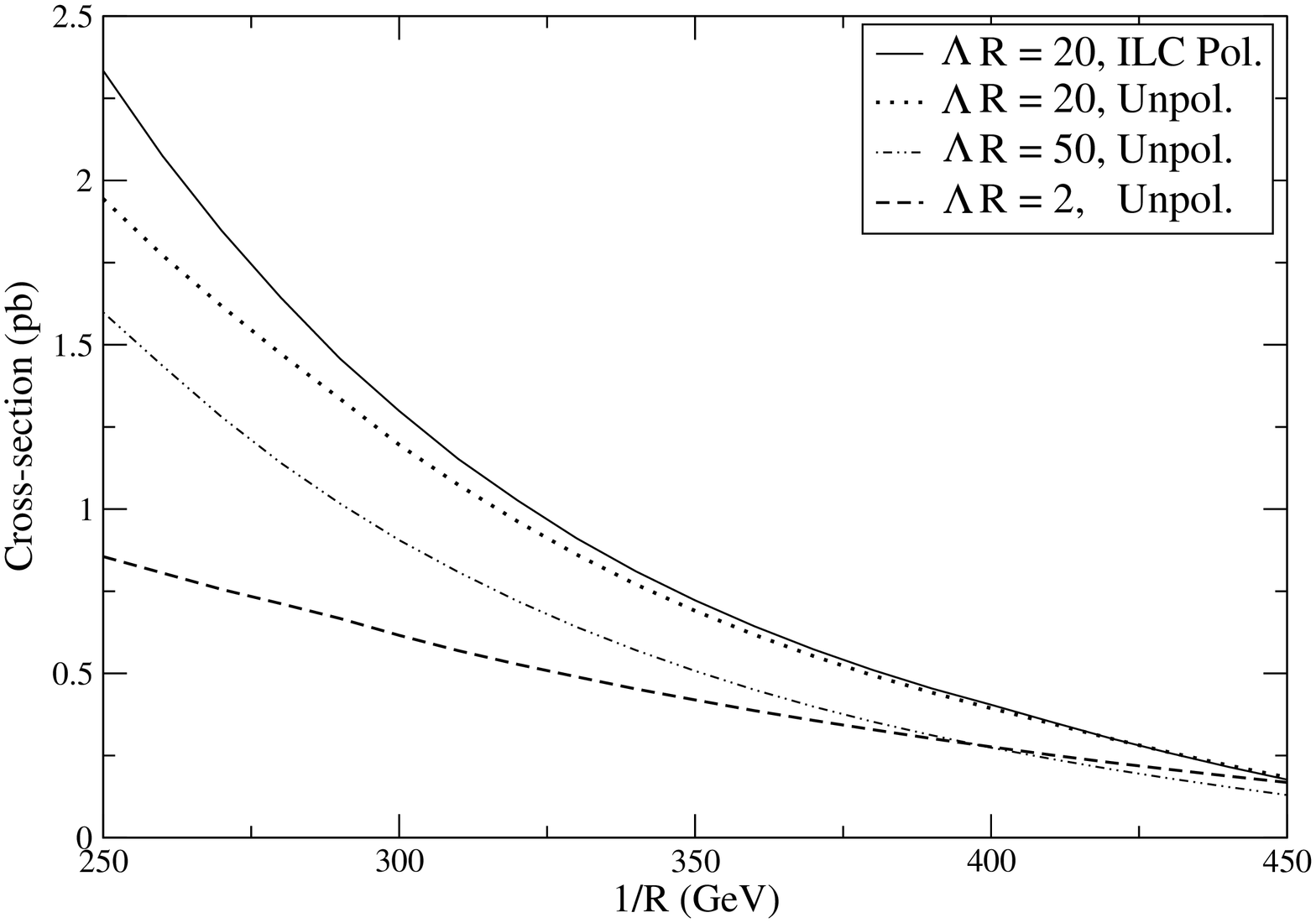}}} 
\hspace{3.3cm}\caption[]{ Cross section versus $1/R$ for the process
$e^+e^-\to e^+e^- + $ missing energy. Plots are shown for unpolarised incident
beams with $\Lambda R =$ 2,20 and 50, and for `optimum' ILC polarisation (80\%
for $e^-$ and 50\% for $e^+$ beams) for $\Lambda R=$ 20. The lower and upper
energy cuts on the final state leptons are set at 0.5 and 20 GeV,
respectively. The angular cuts with respect to the beam axis are set at
$15^\circ$.}  \protect\label{fig1}
\end{figure} 
\begin{figure}
\vspace{-10pt}
\centerline{\hspace{-3.3mm} 
\rotatebox{-90}{\epsfxsize=8cm\epsfbox{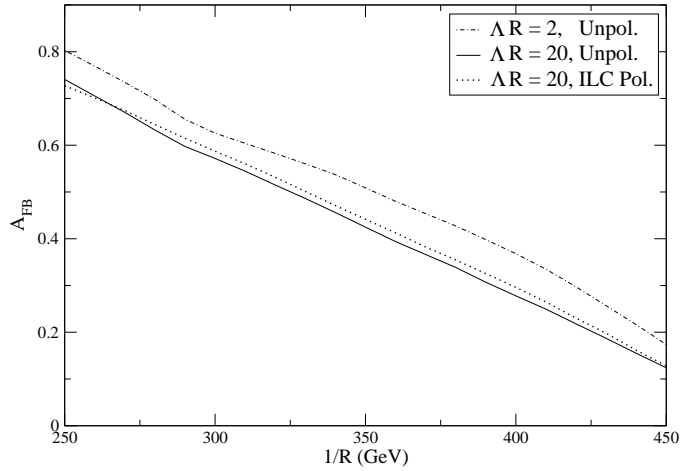}}} 
\hspace{3.3cm}\caption[]{ $A_{FB}$ versus $1/R$ for the same process. Plots
are shown for unpolarised incident beams with $\Lambda R = 2$ and 20, and for
`optimised' ILC polarisation for $\Lambda R=$ 20. The cuts are as in Fig.~1.}
\protect\label{fig2}
\end{figure} 

\end{document}